\documentstyle[preprint,aps]{revtex}
\begin{document}
\draft
\title{Towards dynamical mass calculations}

\author{Ji\v{r}\'{\i} Ho\v{s}ek}
\address{Nuclear Physics Institute, Czech Acad.Sci.,
250 68 \v{R}e\v{z} (Prague), Czech Republic}
\maketitle

\begin{abstract}
$SU(2)_{L} \times U(1)_{Y}$ electroweak gauge model without Higgs sector
is extended by a new vector field $C^{\mu}$ interacting with
leptons and quarks of both chiralities. This interaction is
treated under a dynamical assumption in a self-consistent
approximation. Fermion masses are calculated in terms of new
Abelian hypercharges, and the intermediate boson masses are
calculated in terms of the fermion masses by sum
rules. Self-consistency requires a rich sector of heavy spin-1
collective fermion-antifermion excitations.
\end{abstract}
\pacs{12.15 Ff, 12.60 -i, 14.60 Pq, 14.70 -e}
Experimentally observed mass spectrum of leptons and quarks
ranges from the vanishing or the vanishingly small $(m_{\nu} < O(1
eV))$ neutrino masses over the MeV range of the electron, and u,d
quark masses up to 175 GeV of the recently observed top-quark
mass. Theoretical understanding of such a sparse, wide and
irregular fermion mass spectrum is completely missing. Its
successful phenomenological description in terms of the
independently renormalized Yukawa couplings of fermions to a
condensing elementary Higgs field provides, however, a hope
[1] for the existence of an underlying microscopic dynamics in
which the fermion mass ratios are the calculable numbers.
Alternatively, the problem of understanding the fermion mass
spectrum is shifted towards the very high mass scales in which
the Higgs sector is theoretically likely [2].

We suggest to extend the standard $SU(2)_{L} \times U(1)_{Y}$
gauge-invariant Lagrangian of three massless families f of
leptons and quarks (neutrino singlets $\nu_{fR}$ are added for
completeness) by a vector boson $C^{\mu}$ with the mass $m_{C}$
interacting with all fermions of both chiralities with the
coupling constant h [3].

With a dynamical assumption on the nonperturbative behavior of
the running "Abelian" charge $h^{2}(p)$ we calculate the huge
fermion mass ratios naturally in terms of the new hypercharges
associated with different fermions. It is then suggestive to
call these hypercharges the heaviness. Being the pure numbers,
all of the same order of magnitude they can be fixed (quantized)
by embedding the model into a GUT group at a scale
$M\approx 10^{15}GeV$. This we assume but do not specify the particular
group structure. The intermediate boson masses $m_{W}$ and
$m_{Z}$ are calculated in terms of the fermion masses by sum
rules as a consequence of the $SU(2) \times U(1)$ Ward identities.

The dynamical assumption is that the strong C-boson-fermion interaction
generates specific spin 1 fermion-antifermion composites (collective
excitations) with calculable couplings to the C-boson. These new
interactions then modify the behavior of the running "Abelian" charge
$h^{2}(p)$ to the phenomenologically desirable "walking" form.

Replacement of the Higgs sector by a simple interaction
characterized above brings, however, a serious problem. It is the
problem of unwanted global symmetries. When the fermion masses
are dynamically generated, most of these symmetries are
spontaneously broken and there is nobody except the weakly
coupled W and Z bosons ready to "eat" the corresponding massless
fermion-antifermion Nambu-Goldstone (NG) bosons. The most economic
solution to this problem is to assume that the verified NG
eaters i.e., the massless spin 1 fermion-antifermion composites
are generated by the same nonperturbative dynamics which
generates the NG bosons.
Hence, the role of the assumed dynamically generated spin 1 composites
[4,6] is in fact
3-fold: (i) They absorb the unwanted NG bosons and acquire the
masses
$m^{2}_{V}\sim \alpha^{-1}\pi m^{2}_{W}$.
(ii) They contribute to the C-boson vacuum
polarization and justify a posteriori the dynamical assumption on
the behavior of the running $h^{2}(p)$. (iii) They provide a
distinctive phenomenological signature of the model soon
experimentally testable.

Understanding of the behavior of strongly interacting (Lorentz-invariant)
systems of the quantum fields is largely nonexisting at present. What are
the relevant degrees of freedom (particle-like excitations) in a sensible
perturbative framework is a priori not known, and an underlying effective
field theory appropriate for the considered energy range must look upon
them humbly in accordance with the experimental data. The best example which
we have found of what we suggest here is the Hubbard model [5]:
The (nonrelativistic) system of electrons strongly interacting on the lattice
by a fourfermion interaction can be converted in a number of elaborate mean
field solutions to various phases having the excitations with quite unexpected
properties. In one case, for example, the spin and the charge are carried
separately by the spinons (electrically neutral fermions) and the holons
(electrically charged bosons) at the mean field level,
but there exist important dynamically generated gauge-field fluctuations.

1. Dynamical generation of the fermion masses by chirally
invariant interactions is a genuinely nonperturbative phenomenon
by definition: In perturbation theory the chiral symmetry forces
fermions to stay massless order by order. Operationally the
generation of the fermion mass $m_{f} = \Sigma_{f}(m^{2}_{f})$ amounts
to finding a finite solution (chiral symmetry does not tolerate
the fermion mass counter-terms) of the Schwinger-Dyson equation [7]
for the fermion proper self-energy $\Sigma_{f}(p^{2})$:

\begin{equation}
\Sigma_{f}(p^{2})={3 \over 4}y(f_{L})y(f_{R})h^{2}\int{d^{4}k
\over (2\pi)^{4}} {c[(p-k)^{2}]\over (p-k)^{2}+m^{2}_{C}}
{\Sigma_{f}(k^{2}) \over k^{2}+\Sigma^{2}_{f}(k^{2})}
\end{equation}

In Eq.1 the function $c(p^{2})$ specifies [8] the behavior of
the momentum-dependent charge in the whole momentum range. In
practise it is known explicitly only in a restricted momentum
range where the perturbation theory is justified. Our assumption
of a GUT unification at $p^{2}=M^{2}$ is important:
$c(p^{2}) \sim ln^{-1} (p^{2}/M^{2})$ at $p^{2}\rightarrow \infty$
(asymptotic freedom = AF) implies [9] a finite $\Sigma_{f}(p^{2})$,
since $\Sigma_{f}(p^{2}) \sim p^{-2} (lnp^{2})^{C_{GUT}}$ at $p^{2}
\rightarrow \infty$.

Numerical analyses of Eq.1 reveal that nontrivial solutions
$\Sigma$ start to exist only above certain critical value [10] of
the coupling, $h^{2}_{cr}/4\pi^{2}$ rather large $(\sim 1)$.
Consequently, perturbative form of the Abelian $c(p^{2})$ running
to the Landau pole is highly unlikely. It is truly remarkable
that if $c(p^{2})$ stays essentially constant from 0 to $M^{2}$
and then falls to zero according to AF, the Eq.1 is extremely
sensitive [11] to the details of $c(p^{2})$, and provides a hope
for true calculations of the fermion masses in terms of
heaviness. For an illustration we present the approximate
solution of (1) found for such a case by Akiba and Yanagida [12]:

\begin{equation}
m_{f} \sim \Sigma_{f}(0) \sim M\exp[-\pi/({3 \over 4} y(f_{L})y(f_{R}){h^{2}
\over 4\pi^{2}} - {h^{2}_{cr} \over 4\pi^{2}})^{1/2}]
\end{equation}

Sensitivity to details of $c(p^{2})$ together with the mass
formula (2) lead us to the conclusion that the sparse, wide and
irregular fermion mass spectrum might be understood provided the
prescribed behavior of $c(p^{2})$ is theoretically justified.
Unbearable lightness of calculating the fermion masses in terms
of heaviness using Eq.2 is schematically illustrated by taking
$M = 10^{15} GeV,  h^{2}/4\pi^{2} = 4/3$, $h^{2}_{cr}/4\pi^{2}=1$
and $Y_{f} \equiv {1 \over \pi}[y(f_{L})y(f_{R)}-1]^{1/2}ln 10$ i.e.,
$m_{f}(1/Y_{f}) = M exp(-1/Y_{f})$ : $m_{"\nu"}(24) = 1 eV$,
$m_{"e"}(18) = 1 MeV$, $m_{"d"}(17) = m_{"u"}(17) = 10 MeV$,
$m_{"\mu"}(16) = m_{"s"} (16) = 100 MeV$, $m_{"\tau"}(15) = m_{"c"}
(15) = 1 GeV$, $m_{"b"} (14) = 10 GeV$, $m_{"t"}(13) = 100 GeV$.
We believe these results justify the dynamical assumptions we make.

The arguments in favor of the prescribed form of $c(p^{2})$ are
the following. First, unconventional behavior of the
momentum-dependent charge $c(p^{2})$ can only be due to
additional (nonperturbative) contributions to the C-boson vacuum
polarization tensor. The standard (perturbative) one-loop
contributions of the elementary fermions drive $c(p^{2})$ to the
Landau pole. Second, the new contributions cannot be due to
anything but to the loops of appropriate fermion-antifermion
composites (collective excitations). Third, of a limited arsenal
of the field theory the only known particles the loops of which
can stop the running $c(p^{2})$ to the Landau pole are the
effectively non-Abelian spin 1 fermion-antifermion composites.
Hence, we assume in the following that such massless
spin 1 composites are dynamically generated. We will specify this
new sector below in relation to another reason for its existence.
The spin 1 fermion-antifermion composites in a model
similar to ours were discussed by Lindner and Ross, see Ref.3.

2. That the strongly coupled "Abelian" dynamics in the
nonperturbative regime defined by the prescribed form of
$c(p^{2})$ generates the bosonic collective excitations is
theoretically acceptable and in some cases phenomenologically
most welcome: There is an important instance in our program where
the particular composites are guaranteed by a theorem: If
$\Sigma_{f}(p^{2})$ are found by solving Eq.1 the global chiral
symmetries of the original Lagrangian are spontaneously broken
down to unbroken [13]
$U(1)_{B} \times U(1)_{L_{e}} \times U(1)_{L_{\mu}} \times
U(1)_{L_{\tau}} \times U(1)_{em}$. Since the original global symmetry is
$(U(2)_{L})^{3} \times (U(1)_{R})^{3} \times (U(1)_{R})^{3}$ for quarks and the
same global symmetry for leptons, there are 31 massless
fermion-antifermion composite Nambu-Goldstone bosons coupled to
fermions.

Three of them are needed as we now show. One $SU(2)_{L} \times U(1)_{Y}$
subgroup of the large global symmetry is in fact gauged, i.e.,
there are 4 perturbatively massless gauge
bosons weakly coupled to the fermions. Coupled to the
fermions are, however, also the massless NG bosons. For the
gauged part $SU(2)_{L} \times U(1)_{Y}$ we will employ the corresponding
Ward-Takahashi identities. Being the consequence of the symmetry
of the Lagrangian rather than of the symmetry of the ground
state, they should remain valid regardless of whether the fermion
masses had spontaneously broken the chiral symmetry or not.

The full fermion-W vertex (and analogously the fermion-Z vertex)
allows to calculate the effective W-NG boson vertex (see, e.g.
Ref.3):
\begin{eqnarray}
iq^{\alpha}{g \over 2  \sqrt {2}} N^{1/2}&=&
 \sum_{fermions} \int {d^{4}p \over (2\pi)^{4}}  \\
&\times & Tr {g \over 2
\sqrt {2}} \gamma^{\alpha}(1- \gamma_{5}) S_{u}(p){1 \over
N^{1/2}}
[(1 - \gamma_{5}) m_{d} - (1 + \gamma_{5}) m_{u}] S_{d}(p+q)
\nonumber
\end{eqnarray}
In evaluating the finite integral (3) with $S_f(p)=(
p\hspace{-4pt}\slash\hspace{3pt}+\Sigma_{f}(p^{2}))/(p^{2}-m^{2}_{f})$ (f = u, d where $u
\equiv $ upper, $d \equiv $ down fermion in a doublet) we use for
definiteness and a first orientation the Pagels-Stokar
model [14]
of $\Sigma_{f}(p^{2}) = m_{f}M^{2}/(p^{2}+M^{2})$ consistent within
the log accuracy with
$\Sigma_{f}(p^{2})\sim p^{-2}(ln p^{2})^{C_{GUT}}$ dictated by the
prescribed form of $c(p^{2})$ in (1). As a result, the vertex (3)
gives rise to an important new tree-level contribution to the
longitudinal part of the vacuum polarization $\Pi^{W}_{\mu\nu}(q)$
of the W boson (analogously for the Z boson). Residue at the
massless pole [15] of $\Pi^{W}_{\mu\nu}$ equals $m^{2}_{W}$
(analogously for $m^{2}_{Z}$):

\begin{equation}
m^{2}_{W} = {1 \over 4}g^{2}\sum n_{c} [m^{2}_{u}I_{u;d} +
m^{2}_{d}I_{d;u}]\\
\end{equation}
\begin{equation}
m^{2}_{Z} = {1 \over 4}(g^{2}+g'^{2})\sum n_{c}
[m^{2}_{u}I_{u;u} + m^{2}_{d}I_{d;d}]
\end{equation}
In (4,5) $n_{c}=1$ for leptons and $n_{c} = 3$ for quarks, and the
quantities I are given as

\begin{displaymath}
I_{u;d} = {1 \over 2\pi^{2}} \int^{1}_{0} x ln
{(M^{2}-m^{2}_{d}) x + m^{2}_{d} \over (m^{2}_{u} -
m^{2}_{d}) x + m^{2}_{d}} dx
\end{displaymath}
The sum rules (4,5) are saturated essentially by the top-quark
mass. In contrast to the canonical Higgs case, there is no
genuine weak-interaction mass scale in this approach. As the
formulas (4,5) suggest, the eaten NG bosons are to be viewed as the
particular combinations of many components:

\begin{eqnarray}
\mid\pi^{+}> &=& {1 \over N^{1/2}} \sum
[m_{u}I^{1/2}_{u;d}\mid\bar{u}(1 - \gamma_{5})d> - m_{d}I^{1/2}_{d;u}
\mid\bar{u}(1 + \gamma_{5})d>]\nonumber \\
\mid\pi^{-}> &=& \mid\pi^{+}>^{+}\nonumber \\
\mid\pi^{0}> &=& {1 \over N^{1/2}} \sum
[m_{u}I^{1/2}_{u;u}\mid\bar{u}\gamma_{5}u> -
m_{d}I^{1/2}_{d;d}\mid\bar{d}\gamma_{5}d>]
\end{eqnarray}

3. The combinations of states orthogonal to (6) correspond to 28
unwanted NG bosons of the spontaneous symmetry breaking pattern
$[U(2)]^{6} \times [U(1)]^{12}/[U(1)]^{5}$. The couplings of
these composite NG bosons to the fermions are dictated by the
Goldstone theorem. If the model pretends not to be in a serious
conflict with the experimental facts, these states must not
appear in its physical spectrum. For this reason we specify the
dynamical assumption already made: the same UV stable dynamics
which produces the NG bosons produces in the same channels also
the massless spin 1 fermion-antifermion composites [5] $V^{\mu}$
(i.e., there is a hidden gauge symmetry [4,16]).

Since the origin of the masses of $V^{\mu}$ is the same as of
$m_{W}$ and $m_{Z}$, we take the formulas (4,5) and rescale $g,g'$ by
the large effective V-fermion couplings $f^{2}_{V}/4\pi^{2}\cong
1$. This then leads to a rough estimate that $m^{2}_{V}\cong
\alpha^{-1}\pi m^{2}_{W}$, and it clearly amounts to "carrying the
own skin to the market": The present experimental limits on
the masses of the extra W and Z vector bosons are almost in the
same range: $m_{W'} > 720 GeV$ [17] and $m_{Z'} > 505 GeV$  [18].

4. The effectively non-Abelian dynamically generated vector
composites $V^{\mu}$ make the assumption on the behavior of
$c(p^{2})$ self-consistent: The finite fermion-loop diagrams of
the type examplified in Fig.1 proportional to and vanishing with
the dynamically generated fermion masses give rise (with some
adjustment) to the vertices corresponding to the interaction
Lagrangian

\begin{eqnarray}
{\cal L}^{eff}_{int}& = &- ihC_{h} [\partial_{\mu}V^{+}_{\nu}
(C^{\nu}V^{\mu} - C^{\mu}V^{\nu}) -
\partial_{\mu}V_{\nu}(C^{\nu}V^{\mu+} -
C^{\mu}V^{\nu+})]+\nonumber \\
& & (hC_{h})^{2} [C_{\mu}C^{\mu}V^{+}_{\nu}V^{\nu} -
C_{\mu}V^{\mu}C^{\nu}V^{+}_{\nu}]-\nonumber \\
& & ihC_{h}(\partial_{\mu}C_{\nu} - \partial_{\nu}C_{\mu})V^{\mu+}V^{\nu}
\end{eqnarray}
where $C_{h}$ is a calculable factor. The divergent fermion loops
of Fig.2 not proportional to the fermion masses give rise to the
V-boson kinetic term by virtue of imposing the compositeness
condition Z=0  [4,6] upon the V-boson wave function
renormalization.

Since the interaction (7) is known to produce an antiscreening
[19] contribution to the C-boson vacuum polarization, the
resulting $c(p^{2})$ should walk [20]. For $C_{h}$ fixed as to
cancel the fermion contribution and without GUT unification
$c(p^{2})$ would go to a non-trivial fixed point from above
[21].

Within an effective field theory we have attempted at identifying
the basic dynamics underlying the extremely successful Higgs
phenomenology: (i) It allows for calculating the mass ratios.
(ii) It is economic in the sense of the number of free parameters
in the primary Lagrangian. (iii) It does not shift the solution
of the problem of the mass generation to the far future. In fact
the fermion masses are put on the same footing with the fermion
electric charges, both being fixed by their corresponding Abelian
hypercharges. (iv) It provides distinctive experimental
predictions in the form of the new vector bosons V with masses in
the O(10TeV) range. These collective excitations having the couplings
to quarks should manifest themselves by their propagator effects
if not as resonances both at the Tevatron and HERA. It is perhaps
legitimate to speculate nowadays that the extra events observed
at very high $Q^{2}$ in the ep collisions at HERA are namely due
to them. (v) Its detailed description clearly demands
further elaboration.

Field-theoretically, there is no a priori way of knowing whether our
dynamical assumption on the nonperturbative behavior of the Abelian
$\beta$ - function is justified or not. We cannot refrain from pointing
out that the concept of the dynamically generated gauge particles which is
crucial in our approach is seriously considered in the effective
field theory description of the high-$T_{c}$
superconductors [5].
This analogy also nicely illustrates an uneasy and ambiguous task
of finding the correct microscopic dynamics to a given (say, Higgs)
phenomenological one: While both the standard and high-$T_{c}$
superconductors are equally well parametrized by the same Ginzburg-
Landau-Higgs phenomenological theory, their microscopic dynamics are
apparently quite distinct: While the standard superconductors are
definitely governed by the microscopic BCS dynamics, understanding
of the microscopic dynamics of the high-$T_{c}$ superconductors is
still far from  complete.

The work was supported by the grant AV\v{C}R 148402. A short-term
visit of the author at CERN-TH where the present paper was
completed was generously supported by the ATLAS budget of the
Committee for CR-CERN Cooperation.

\begin{figure}
\caption{The UV finite Feynman diagrams giving rise to the effective
vertices of the Lagrangian (7).}
\end{figure}
\begin{figure}
\caption{The divergent Feynman diagram giving rise to the V-boson
kinetic term [4,20].}
\end{figure}


\begin{thebibliography}{99}
\bibitem{1} Our belief is based upon the common property of good
phenomenological theories: Their parameters are calculable within
underlying dynamics considered microscopic at the next typical
energy scale. For us the guiding, relevant and stimulating are
the interrelations between various theories of superconductivity.
\bibitem{2} The sector of elementary Higgs fields is made likely
within SUSY.
\bibitem{3} Some aspects of the dynamical mass generation were
elaborated within the present Lagrangian in: J. Ho\v{s}ek,
{\it Phys. Rev.} {\bf D36} (1987) 2093. A closely related model
is: M. Lindner and D. Ross, {\it Nucl. Phys.} {\bf B370} (1992)
30. Heuristic elaboration of the dynamical mass generation
amounts to dealing with an effective fourfermion interaction:
J. Ho\v{s}ek, preprint CERN-TH.4104/85 (unpublished);
W. A. Bardeen, C. T. Hill and M. Lindner, {\it Phys. Rev.}
{\bf D41} (1990) 1647.
\bibitem{4} M. Bando, Y. Taniguchi and S. Tanimura, preprint KU-AMP
96014 (hep-th/9610244) and references quoted therein.
\bibitem{5} G. Baskaran and P. W. Anderson, {\it Phys. Rev.} {\bf
B37} (1988) 580; P. W. Wiegmann, {\it Phys. Rev. Lett.} {\bf 60}
(1988) 821; L. B. Ioffe and A. I. Larkin, {\it Phys. Rev.} {\bf B39}
(1989) 8988; J. B. Marston and I. Affleck, {\it Phys. Rev.} {\bf
B39} (1989) 11538; E. Fradkin, Field Theories of Condensed Matter
Systems, Addison-Wesley, Redwood City, 1991; J. Polchinski,
preprint UTTG-09-93.
\bibitem{6}  K. Akama, {\it Phys. Rev. Lett.} {\bf 76} (1996) 184
and references quoted therein; K. Akama and T. Hattori, {\it Phys. Lett.}
{\bf B392} (1997) 383.
\bibitem{7} H. Pagels, {\it Phys. Rev.} {\bf D21} (1980) 2336.
\bibitem{8} It also specifies the proper gauge: H.Georgi,
E. S. Simmons and A. Cohen, {\it Phys. Lett.} {\bf B236} (1990)
183.
\bibitem{9} K. Lane, {\it Phys. Rev.} {\bf D10} (1974) 2605; H. D.
Politzer, {\it Nucl. Phys.} {\bf B117} (1976) 397.
\bibitem{10} T. Maskawa and H. Nakajima, {\it Progr. Theor. Phys.} {\bf
52} (1974) 1326.
\bibitem{11} B. Holdom, {\it Phys. Lett.} {\bf 150B} (1985) 301.
\bibitem{12} T. Akiba and T. Yanagida, {\it Phys. Lett.} {\bf
169B} (1986) 432; see also: C. N. Leung, S. T. Love and W. A.
Bardeen, {\it Nucl. Phys.} {\bf B273} (1986) 649; V. A. Miransky,
{\it Nuovo Cim.} {\bf 90A} (1985) 149. The mass $m_{C}$ is put to
zero in Refs. 8,9. Phenomenologically $m_{C}> 10^{5}GeV$
(restriction due to FCNC), which is still much smaller than
$M\cong 10^{15}GeV$.
\bibitem{13} We adopt the standard solution of the strong $U(1)_{A}$
problem.
\bibitem{14} H. Pagels and S. Stokar, {\it Phys. Rev.} {\bf D20}
(1979) 2947.
\bibitem{15} J. Schwinger, {\it Phys. Rev.} {\bf 128} (1962)
2425; P. W. Anderson, {\it Phys. Rev.} {\bf 130} (1963) 439.
\bibitem{16} M. Suzuki, {\it Phys. Rev.} {\bf D37} (1988) 210;
A. Cohen, H. Georgi and E. H. Simmons, {\it Phys. Rev.} {\bf D38}
(1988) 405; D. A. Kosower, {\it Phys. Rev.} {\it D48} (1993) 1288.
\bibitem{17}DO Collaboration, S. Abachi et al., {\it Phys. Rev.
Lett.} {\bf 76} (1996) 3271.
\bibitem{18} CDF Collaboration, F. Abe et al., {\it Phys. Rev.}
{\bf D51} (1995) 949.
\bibitem{19} N. K. Nielsen, {\it Am. J. Phys.} {\bf 49} (1981)
1171; D. J. Gross and F. Wilczek, {\it Phys. Rev. Lett.} {\bf 30}
(1973) 1343; H. D. Politzer, {\it Phys. Rev. Lett.} {\bf 30}
(1973) 1346.
\bibitem{20} B. Holdom, {\it Phys. Rev.} {\bf D24} (1981) 1441;
T. Appelquist and L. C. R. Wijewardhana, {\it Phys. Rev.} {\bf
D36} (1987) 586.
\bibitem{21} T. Appelquist, J. Terning and L. C. R. Wijewardhana,
{\it Phys. Rev. Lett.} {\bf 77} (1996) 1214.
\end{thebibliography}
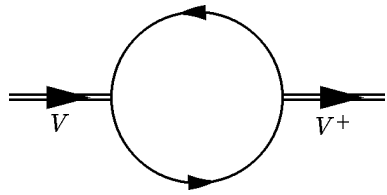
\end{document}